# Spin glass state in layered compound MnSb$_2$Te$_4$

Hao Li[1,6], Yaoxin Li[3], Yu-Kun Lian[4], Weiwei Xie[5], Ling Chen[4], Jinsong Zhang[3,7], Yang Wu[2,6,*], Shoushan Fan[3,6]

[1]*School of Materials Science and Engineering, Tsinghua University, Beijing, 100084, P. R. China*
[2]*Department of Mechanical Engineering, Tsinghua University, Beijing, 100084, P. R. China*
[3]*State Key Laboratory of Low Dimensional Quantum Physics and Department of Physics, Tsinghua University, Beijing 100084, P. R. China*
[4]*Beijing Key Laboratory of Energy Conversion and Storage Materials, College of Chemistry, Beijing Normal University, Beijing 100875, P. R. China*
[5]*Department of Chemistry and Chemical Biology, Rutgers university, Piscataway, New Jersey, 08854, USA*
[6]*Tsinghua-Foxconn Nanotechnology Research Center, Tsinghua University, Beijing 100084, P. R. China*
[7]*Frontier Science Center for Quantum Information, Beijing, P. R. China*

As a sister compound and isostructural of MnBi$_2$Te$_4$, the high quality MnSb$_2$Te$_4$ single crystals are grown via solid-state reaction where prolonged annealing and narrow temperature window play critical roles on account of its thermal metastability. X-ray diffraction analysis on MnSb$_2$Te$_4$ single crystals reveals pronounced cation intermixing, 28.9(7)% Sb antisite defects on the Mn (3a) site and 19.3(6)% Mn antisite defects on the Sb (6c) site, compared with MnBi$_2$Te$_4$. Unlike antiferromagnetic (AFM) nature MnBi$_2$Te$_4$, MnSb$_2$Te$_4$ contains magnetic and antiferromagnetic competition and exhibits a spin glass (SG) state below 24 K. Its magnetic hysteresis, anisotropy, and relaxation process are investigated in detail with DC and AC magnetization measurements. Moreover, anomalous Hall effect as a p-type conductor is demonstrated through transport measurements. This work grants MnSb$_2$Te$_4$ a possible access to the future exploration of exotic quantum physics by removing the odd/even layer number restraint in intrinsic AFM MnBi$_2$Te$_4$-family materials as a result of the crossover between its magnetism and potential topology in the Sb-Te layer.

# I. Introduction

Pursuit of emerging topological materials that cultivate exotic quantum phenomena, for instance, dissipationless electronics transport and topological quantum computation, has been a research frontier in condensed matter physics and material science[1-3]. In this regard, magnetic topological quantum materials (MTQMs) can be one of the most fascinating examples, owing to its interplay of nontrivial topology and magnetism that could give rise to numerous transport phenomena, such as the quantum anomalous Hall (QAH) effect and the axion insulator state[3-11]. One of the important examples of MTQMs is the Cr-doped $(Bi_xSb_{1-x})_2Te_3$ thin film fabricated in a molecular epitaxy pross that hosts the QAH effect[7]. However, subsequent studies on magnetic topological insulators (TIs) revealed that doping with magnetic impurity was a fairly challenging strategy in realizing homogeneous distribution in materials, restricting significant improvement in observing quantized conductance in high temperatures and hindering the experimental progress and further application[8-13]. Therefore, intrinsic MTQMs with naturally ordered magnetic-topological interplays could serve as a cornerstone for the future discovery of numerous emerging topological properties in materials[14].

Recently, $MnBi_2Te_4$, a representative ternary chalcogenide with a general formula of $MB_2T_4$ (M = transition-metal or rare-earth element, B = Bi or Sb, T = Te, Se, or S), has been identified to be the first intrinsic magnetic TI by both theoretical and experimental efforts[14-32]. As a van der Waals layered compound with a tetradymite-derived structure, $MnBi_2Te_4$ (space group $R\bar{3}m$) consists of Te-Bi-Te-Mn-Te-Bi-Te septuple layers (SLs) stacking in an ABC sequence. Appealingly, magnetic order and nontrivial topology inherently coexist in $MnBi_2Te_4$ and it belongs to MTQMs. On the one hand, each $Mn^{2+}$ offers a magnetic moment of 5 $\mu_B$ according to Hund's rule in high-spin configuration. The intralayer exchange coupling between Mn-Mn is ferromagnetic (FM) with an out-of-plane easy axis, while the interlayer coupling between neighboring SLs is antiferromagnetic (AFM), corresponding to an A-type AFM order. On the other hand, the Bi-Te layers could generate topological states by their strong spin-orbit coupling. The magnetic and topological property of $MnBi_2Te_4$ together generate a gapped surface state by symmetry-breaking on the (0001) surface, confirmed experimentally on the high-quality $MnBi_2Te_4$ single crystal[21]. Moreover, $MnBi_2Te_4$ revealed layer number dependent topological phase evolution, where the long-sought axion insulator phase and QAH effect have been realized in mechanically exfoliated six-SL and five-SL $MnBi_2Te_4$ films, respectively[31,32]. The thriving research on $MnBi_2Te_4$ not only contributes to the emerging topological quantum states, but also motivates the synthesis of potential MTQMs. The great variety of unexplored $MB_2T_4$ family is highly possible to generate tremendous novel topological quantum phases and thus arouse further interests in quantum physics.

Herein, we report the growth of $MnSb_2Te_4$ single crystals, a typical member of the $MB_2T_4$-family materials, and establish its property in terms of thorough characterizations of the structure, composition, thermal and magnetic analysis. Interestingly, $MnSb_2Te_4$ single crystal exhibits spin glass (SG) behaviors below $T_{SG} \sim 24$ K, different from the predicted and experimentally revealed A-type AFM state as $MnBi_2Te_4$, or FM interaction reported in polycrystalline $MnSb_2Te_4$[33-35]. In fact, considerable net magnetic moment is observed in $MnSb_2Te_4$ single crystals below its SG transition temperature, which makes it a promising candidate to realize intriguing quantized conductance in zero-magnetic-field, without the restraint of layer number dependence in AFM $MnBi_2Te$[14].

## II. Experimental

MnTe precursor was synthesized by directly heating a stoichiometric mixture of high-purity Mn (99.95%, Alfa Aesar) and Te (99.999%, Aladdin) at 1273 K in a vacuum-sealed silica ampoule for 3 days. The ampoule was quenched in air to obtain MnTe precursor. $Sb_2Te_3$ precursor was synthesized by a solid-state reaction of a stoichiometric mixture of high-purity Sb (99.99%, Alfa Aesar) and Te (99.999%, Aladdin). The reaction mixture was vacuum-sealed in a silica ampoule, and heated to 1073 K and held for 24 h. After being slowly cooled to 863 K at a rate of 0.1 K min$^{-1}$ and held for 48 h, the ampoule was quenched in air to obtain $Sb_2Te_3$ precursor.

Morphology and element analyses were carried out using a FEI NOVA SEM450 scanning electron microscope (SEM) equipped with an energy dispersive X-ray (EDX) detector. X-ray photoelectron spectroscopy measurements were conducted using an Ulvac-Phi Quantera II X-ray photoelectron spectrometer with monochromatic Al Ka radiation (1486.6 eV). The C 1s peak at 284.8 eV is used as the reference. Single crystal X-ray diffraction (SCXRD) data were collected at 298 K on a Bruker D8 Quest diffractometer using monochromatic Mo Ka radiation ($\lambda$ = 0.71073 Å). The absorption correction was done by the multiscan method. The structure was solved by SHELXT and refined by SHELXL in OLEX2[36-38]. Powder X-ray diffraction (PXRD) patterns were collected on a Rigaku D/max-2500/PC X-ray diffractometer, using Cu K$\alpha$ radiation and operating at 40 kV and 200 mA. Raman spectra were collected on a Horiba Jobin Yvon LabRam-HR/VV Spectrometer with a 514 nm laser source and a 1800-line grating. Differential thermal analysis (DTA) and thermogravimetric (TG) analysis were carried out using a NETZSCH STA449 F3 simultaneous thermal analyzer under $N_2$ atmosphere. The $MnSb_2Te_4$ sample was enclosed in an $Al_2O_3$ crucible, and heated from room temperature to 973 K at a rate of 10 K min$^{-1}$ and then cooled to 673 K at a rate of 10 K min$^{-1}$.

Magnetic property measurements were carried out using a magnetic property measurement system (MPMS SQUID VSM Quantum Design). Dynamic (AC) magnetization measurements were performed with an oscillating field, which can be written as, $H(t) = H_{DC} + H_{AC}\cos(\omega t)$, where $t$ is time, $\omega$ is angular frequency, $H_{DC}$ is the magnitude of DC field, and $H_{AC}$ is the amplitude of oscillating field[39,40]. The magnetization response of a measured material can be expressed as, $M(t) = M_{DC} + M_{AC}\cos(\omega t − \phi)$, where $\phi$ is a phase shift due to magnetic relaxation. AC susceptibility can be defined as, $\chi = \chi' − i\chi''$, in which, the real term, $\chi' = M_{AC}\cos(\phi)/H_{DC}$, keeps in-phase with the AC field, and the imaginary term, $\chi'' = M_{AC}\sin(\phi)/H_{DC}$, is out-of-phase with the AC field and is a measure of the energy losses due to magnetic relaxation.

For a single relaxation time ($\tau$), AC susceptibility can be described by the Debye model as, $\chi(\omega) = \chi + (\chi_0 - \chi_\infty)/(1 + i\omega\tau)$, where $\chi_0$ and $\chi_\infty$ are the susceptibility at zero frequency and extremely large frequency, respectively. In actual magnetic materials, there is a distribution for $\tau$, and a generalized Debye model, $\chi(\omega) = \chi + (\chi_0 - \chi_\infty)/(1 + i\omega\tau)^{(1 - \alpha)}$, where $\alpha$ is the distribution parameter for $\tau$, is used to describe the $\chi''$ - $\chi'$ plot (Cole-Cole plot). The derived parameters $\alpha$ and $\tau$ provide specific features of a magnetic material[39-42].

Transport property measurements were carried out using a cryostat (Oxford Instruments) with a base temperature of ~1.6 K and a magnetic field up to 2 T. The longitudinal and Hall voltages were detected simultaneously by using Stanford Research Instrument SR830 lock-in amplifiers with an AC current generated with a Keithley 6221 current source. The $MnSb_2Te_4$ single crystal sample with a thickness of ~35 μm was obtained by mechanical exfoliation from a raw $MnSb_2Te_4$ single crystal by Scotch tape.

## III. Results and discussion

### A. Crystal growth

Learning from the thermal stability and synthetic strategy of MnBi$_2$Te$_4$ crystal[29], we adopted the similar growth strategy with extended annealing for the crystal growth of MnSb$_2$Te$_4$, predicted to be isostructural to MnBi$_2$Te$_4$. Firstly, Sb$_2$Te$_3$ and MnTe binary used as precursors were prepared through direct reactions of high-purity Sb and Te, and Mn and Te in stoichiometric mixtures, respectively (Experimental Synthesis Section). Hereafter, high-quality MnSb$_2$Te$_4$ single crystals were well grown from a 1 : 1 mixture of Sb$_2$Te$_3$ and MnTe. The vacuum-sealed sample was heated to 1173 K in 9 hours and slowly cooled to 893 K at a rate of 1 K min$^{-1}$, followed by prolonged annealing at 893 K for at least 14 days. Finally, the sample was quenched in air and cracked to pick shiny crystals by cautious selection. To remove impurities, coarse parts of crystals were carefully cut off using a scalpel or cleaved off by scotch tape with the aid of an optical microscope. The growth route can successfully afford millimeter-size MnSb$_2$Te$_4$ single crystals (Figs. 1(a) and 2(b)).

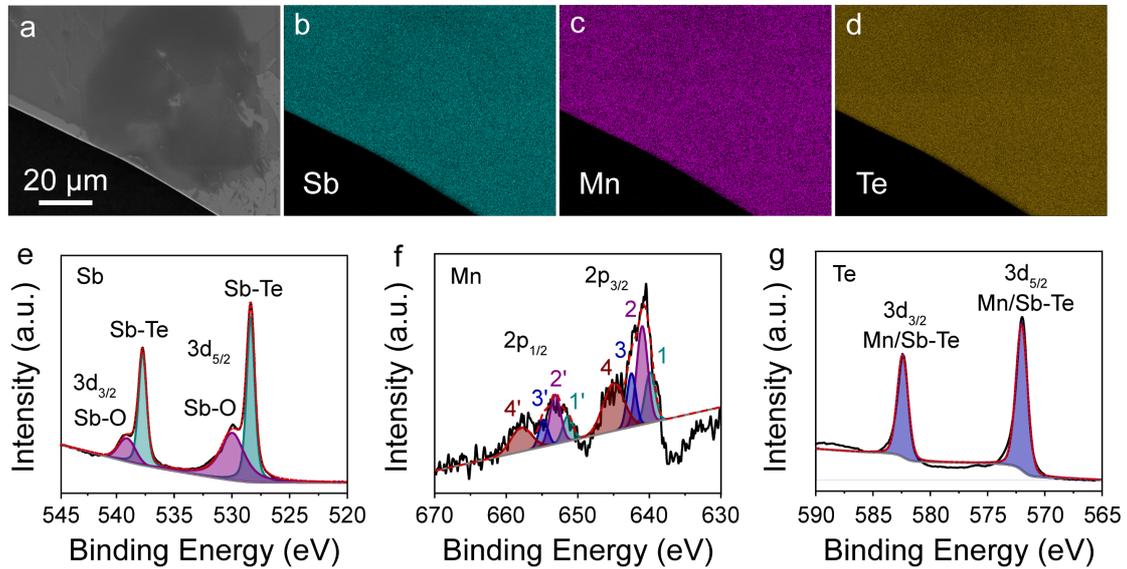

FIG. 1. Morphology and elemental analyses of exfoliated MnSb$_2$Te$_4$ single crystals. (a) SEM and (b-d) EDX elemental mapping images (Sb, Mn and Te). (e) Sb 3d, (f) Mn 2p and (g) Te 3d high-resolution XPS spectra.

### B. Composition and oxidation state analysis

To get rid of the influence from ambient contamination, we prepared fresh surface of MnSb$_2$Te$_4$ crystals by cleaving them with a scotch tape. The scanning electron microscopy (SEM) image of exfoliated MnSb$_2$Te$_4$ crystal shows asmooth surface (Fig. 1(a)). Distinctive edges between MnSb$_2$Te$_4$ layers of different thickness can be observed, confirming it a van der Waals layered compound. Therefore, it is reasonable to obtain high-quality few-SL MnSb$_2$Te$_4$ flakes via mechanical exfoliation. Energy dispersive X-ray (EDX) spectroscopy was further adopted to analyze the composition. The EDX spectrum (see Fig. S2a in the Supplemental Material) displays no peaks other than Sb, Mn, Te and C (from carbon tape). In addition, corresponding EDX elemental mapping images in Figs. 1(b)-(d) clearly show a uniform elemental distribution of Sb, Mn and Te.

Quantification based on the EDX spectrum using Sb L, Te L and Mn K edges has given the composition as Mn 15.6, Sb 26.9, Te 57.4 at%, corresponding to the chemical formula of $Mn_{1.1}Sb_{1.9}Te_{4.0}$.

We carried out X-ray photoelectron spectroscopy (XPS) measurement on the exfoliated fresh surface of $MnSb_2Te_4$ crystal to determine the oxidation states. The XPS survey spectrum (see Fig. S2b in the Supplemental Material) consolidates the constituent elements as Sb, Mn and Te. In the high-resolution Sb 3d spectrum in Fig. 1(e), two major peaks at 537.7 eV (Sb $3d_{3/2}$) and 528.3 eV (Sb $3d_{5/2}$) can be assigned to $Sb^{2+}$ in Sb-Te bonds[43,44]. Two shoulder peaks at 539.0 eV and 530.0 eV corresponding to Sb-O bonds imply the slight oxidation due to the short exposure to air[43,44]. Figure 1(f) shows the deconvoluted Mn 2p spectrum that consists of two sets of four sub peaks (Mn1('), Mn2('), Mn3('), and Mn4(')). Mn1 and Mn1' peaks located at 639.9 eV and 651.4 eV could be attributed to broken Mn-Te bonds by Te deficiency. The strongest Mn2 and Mn2' peaks at 641.0 eV and 653.1 eV define the major oxidation state of Mn as 2+. Mn3 and Mn3' peaks at 642.5 eV and 654.8 eV, and Mn4 and Mn4' peaks at 645.0 eV and 657.9 eV are satellite peaks of Mn1 and Mn1', and Mn2 and Mn2', respectively, which could result from the charge transfer between the unfilled 3d shell of Mn and the outer shell of Te in the photoelectron process, similar to $MnBi_2Te_4$[45-48]. For the Te 3d spectrum in Fig. 1(g), it contains two peaks at 572.0 eV (Te $3d_{5/2}$) and 582.4 eV (Te $3d_{3/2}$) that can be assigned to $Te^{2-}$ in Mn-Te and Sb-Te (Mn/Sb-Te) bonds[43,44]. To gain insight into its oxidation behavior, the $MnSb_2Te_4$ crystal was exposed to air for more than a week and investigated using XPS. As shown in Fig. S3 in the Supplemental Material, the Sb-O and Te-O signals turn out to be evidently higher than Sb-Te and Mn/Sb-Te peaks in the high-resolution Sb 3d and Te 3d spectra, indicating serious surface oxidationg that t $MnSb_2Te_4$ suffers in air condition. On this account, it is necessary to keep $MnSb_2Te_4$ samples, especially the few-layer one, in an inert atmosphere to avoid the damage caused by oxidation.

### C. Crystal Structure and Refinement

The crystal structure of the as-grown $MnSb_2Te_4$ crystal was elucidated via SCXRD measurement. As an isostructural compound to $MnBi_2Te_4$, $MnSb_2Te_4$ also crystallizes in the centrosymmetric space group $R\bar{3}m$ (no. 166) featuring the Te-Sb-Te-Mn-Te-Sb-Te SLs stacking in an ABC sequence along the c-axis (see Fig. 2(a)). Table I lists the lattice parameters of $MnSb_2Te_4$ are $a = 4.2284(5)$ Å and $c = 40.705(1)$ Å), slightly smaller than those of $MnBi_2Te_4$. This is reasonable taking the smaller ionic radius of Sb than that of Bi into account. Besides, cation antisite disorder is ubiquitous in $MB_2T_4$-type compounds[15,35,49]. For this reason, antisite mixing of Sb and Mn was taken into account in the structural refinement by allowing Sb (6c) to occupy Mn (3a) site and vice versa. Te vacancies were only allowed in Te2 position that occupies the outermost layer in the Te-Sb-Te-Mn-Te-Sb-Te SLs, because those Te atoms are easier to lose considering the weak van der Waals bonding in $MnSb_2Te_4$. Charge neutrality was kept in the structure refinement under the restraint of eight positive charges per formula unit by considering $Mn^{II}(Sb^{III})_2(Te^{-II})_4T$ oxidation states according to XPS results.

The refinement generated a nonstoichiometric model consisting of 28.9(7)% Sb occupancy on the Mn (3a) site, 19.3(6)% Mn occupancy on the Sb (6c) site and 2.4(5)% void defects on the Te2 (6c) site. The composition that amounts to $Mn_{1.09(7)}Sb_{1.90(3)}Te_{3.95(1)}$ is in good consistent with the SEM-EDX quantification result of $Mn_{1.1}Sb_{1.9}Te_{4.0}$. The full structural information are shown in Table I-II and Table SI-SII in the Supplemental Material. For the sake of convenience, the formula

MnSb$_2$Te$_4$ instead of Mn$_{1.09(7)}$Sb$_{1.90(3)}$Te$_{3.95(1)}$ is still used throughout the article. For comparison, the cation intermixing is in a much lighter extent in MnBi$_2$Te$_4$ prepared using the same prolonged annealing method, which just comprises 16.(2)% Bi in the Mn (3a) site (see Table SIII-SV in the Supplemental Material). Please note that MnBi$_2$Te$_4$ crystals grown by slow cooling within a narrow range exhibit similar antisite defects on the Mn 3a site (21.5(1)% Bi, 73.6(1)% Mn, and 4.9(1)% voids) and 5.7(1)% of Mn mixing on the Bi (6c) site[15]. Therefore, MnSb$_2$Te$_4$ affords much more pronounced cation antisite disordering than MnBi$_2$Te$_4$, and this may induce new magnetism distinct to AFM coupling.

TABLE I. Crystallographic data of Mn$_{1.09(7)}$Sb$_{1.90(3)}$Te$_{3.95(1)}$, refined with cation intermixing and Te vacancies from SCXRD.

| | |
|---|---|
| empirical formula | Mn$_{1.09(7)}$Sb$_{1.90(3)}$Te$_{3.95(1)}$ |
| formula units | $Z = 3$ |
| crystal system, space group | trigonal, $R\bar{3}m$ (no. 166) |
| lattice parameters | $a = 4.2284(5)$ Å |
| | $c = 40.705(10)$ Å |
| | $V = 630.3(2)$ Å$^3$ |
| temperature | 304.36 K |
| range for data collection; index ranges | $6.006° \leq 2\theta \leq 60.982°$ ($\lambda = 0.71073$ Å) |
| | $-5 \leq h \leq 6, -6 \leq k \leq 6, -57 \leq l \leq 57$ |
| collected reflections | 3406 measured, 286 unique |
| $R$ indexes of merging | $R_{int} = 0.1012, R_\sigma = 0.0364$ |
| data/restraints/parameters | 286/1/15 |
| final $R$ indexes [$I > 2\sigma(I)$] | $R_1 = 0.0526, wR_2 = 0.1346$ |
| Final $R$ indexes (all data) | $R_1 = 0.0564, wR_2 = 0.1386$ |
| goodness-of-fit on $F^2$ | 1.130 |

TABLE II. Crystallographic data of Mn$_{1.09(7)}$Sb$_{1.90(3)}$Te$_{3.95(1)}$, refined with a cation intermixing model from an SCXRD experiment.

| atom | $x$ | $y$ | $z$ | $U_{eq}$ (pm$^2$) |
|---|---|---|---|---|
| Te1 6c | 2/3 | 1/3 | 0.04089(3) | 189(4) |
| Te2 6c | 0 | 0 | 0.13122(3) | 188(4) |
| Sb1 6c | 1/3 | 2/3 | 0.09149(4) | 216(5) |
| Sb2 3a | 0 | 0 | 0 | 220(5) |
| Mn1 6c | 1/3 | 2/3 | 0.09149(4) | 216(5) |
| Mn2 3a | 0 | 0 | 0 | 220(8) |

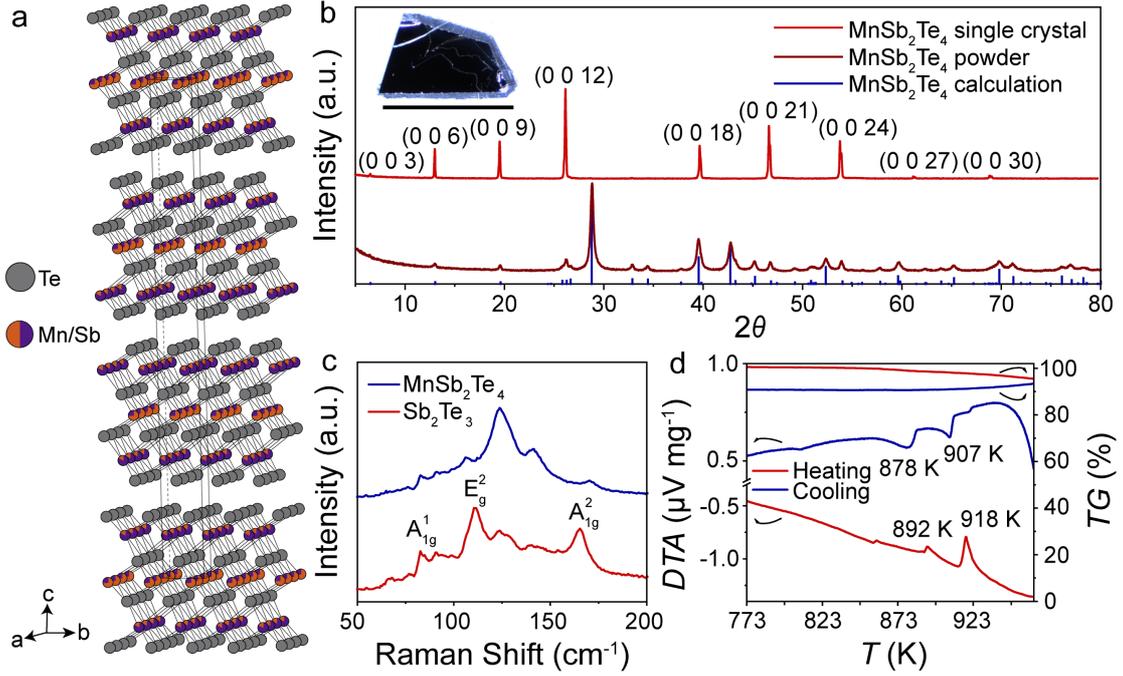

FIG. 2. (a) Crystal structure of the $Mn_{1.09(7)}Sb_{1.90(3)}Te_{3.95(1)}$ model with cation intermixing. The unit cell is marked. (b) PXRD patterns of $MnSb_2Te_4$ single crystal (top) and powder (middle). Inset: An optical image of the as-grown $MnSb_2Te_4$ single crystal, and the scale bar is 2 mm. (c) Raman spectra of $MnSb_2Te_4$ and $Sb_2Te_3$. (d) Differential thermal analysis (DTA) and thermogravimetric (TG) analysis curves of $MnSb_2Te_4$.

The PXRD pattern of raw $MnSb_2Te_4$ single crystal in Fig. 2(a) displays sharp and intense peaks that follow the (0 0 $l$), $l$ = 3n, diffraction rule, indicating the rhombohedral symmetry and a strongly preferred orientation along the [0 0 1] direction. As displayed in Fig. 2(b), the PXRD pattern of ground $MnSb_2Te_4$ powder is in good agreement with the one calculated from the aforementioned structural model of $Mn_{1.09(7)}Sb_{1.90(3)}Te_{3.95(1)}$. No impurity peaks are observed and sharp and intense peaks are displayed, confirming the high purity and crystallinity of the as-grown $MnSb_2Te_4$ single crystal. The Raman spectrum of $MnSb_2Te_4$ crystal in Fig. 2(c) shows an appreciable blue shift and weaker $A_{1g}^1$ and $A_{1g}^2$ peaks associated with $Sb_2Te_3$, which might indicate some impurities that can not be dectected by PXRD[50,51].

### D. Thermal stability

The thermal property of $MnSb_2Te_4$ was revealed using differential thermal analysis (DTA). As shown in Fig. 2(d), the DTA curves display an intense exothermic peak at 918 K on the heating trace and an endothermic peak at 907 K on the cooling trace, corresponding to the melting and crystallization of $MnSb_2Te_4$ respectively. The lower intensity of the endothermic peak at 918 K compared with that of the exothermic peak at 907 K indicates that the crystallization of $MnSb_2Te_4$ is slow and difficult. An exothermic peak at 892 K on the heating trace and an endothermic peak at 878 K on the cooling trace can be ascribed to the melting and crystallization of $Sb_2Te_3$ respectively, revealing that $MnSb_2Te_4$ might be partly decomposed to $Sb_2Te_3$ during the heating. The exothermic peak at 892 K is considerably low while the endothermic peak at 878 K is stronger, suggesting that the decomposition of $MnSb_2Te_4$ proceeds notably faster upon the melting of $Sb_2Te_3$ and $MnSb_2Te_4$.

The weight loss in the thermogravimetric (TG) curve in Fig. 2(d) could arise from the volatilization of Sb or Te. Accordingly, the synthesis of MnSb$_2$Te$_4$ single crystals requires strict control of the annealing temperature, balancing adequate energy for the reacting MnTe and Sb$_2$Te$_3$ and avoiding the severe decomposition of MnSb$_2$Te$_4$.

### E. Magnetic and transport properties

To check the magnetic characteristics of the as-grown MnSb$_2$Te$_4$ single crystal, static (DC) magnetization measurements were performed. Figures 3(a)-3(b) show the temperature dependence of magnetic susceptibility ($\chi$) for the MnSb$_2$Te$_4$ single crystal in out-of-plane (***H*** // c) and in-plane (***H*** // ab) magnetic fields, respectively. As shown in Fig. 3(a), the paramagnetic (PM) regime follows Curie-Weiss law, $\chi(T) = \chi_0 + C/(T - \theta_{CW})$, where $\chi_0$ includes core diamagnetism and Pauli paramagnetism. The Curie-Weiss fitting (70 - 250 K region) yields the effective magnetic moment of $\mu_{eff}$ of 5.4 $\mu_B$ by taking the relationship $C = N_A\mu_{eff}^2/3k_B$ into account, in good consistence with the value for high-spin state of Mn$^{2+}$ (5.92 $\mu_B$). At a small field of 0.01 T, a large bifurcation between zero-field-cooled (ZFC) and field-cooled (FC) curves occurs below $T_{SG}$ ~ 24 K in both ***H*** // c and ***H*** // ab, which is a hallmark for spin glass (SG) state[41,52-55]. At higher fields of 0.1 T for ***H*** // c and 2 T for ***H*** // ab, ZFC and FC curves coincides, indicating it enters a ferromagnetic (FM) state. The larger magnetic susceptibility in the low temperature region (5-30 K) in ***H*** // c than that in ***H*** // ab suggests a strong magnetic anisotropy and the magnetization-easy axis along c direction in MnSb$_2$Te$_4$. The magnetic anisotropy is further revealed in the field dependence of magnetization in Figs. 3(c)-3(d). The *M-H* curves in ***H*** // c (see Fig. 3(c)) display obvious hysteresis loop below 20 K featured by a remanent magnetization of about 0.001 $\mu_B$/Mn and a coercive field of about 0.01 T at 5 K, consistent with the SG state below $T_{SG}$ ~ 24 K. While the linear *M-H* curves in ***H*** // c above 30 K indicate the PM state. The *M-H* curves at low fields in ***H*** // ab (see Fig. 3(d)) display linear dependence with no hysteresis, suggesting that the magnetization along ab plane is merely due to the field-bent magnetic moments originally sited along c direction.

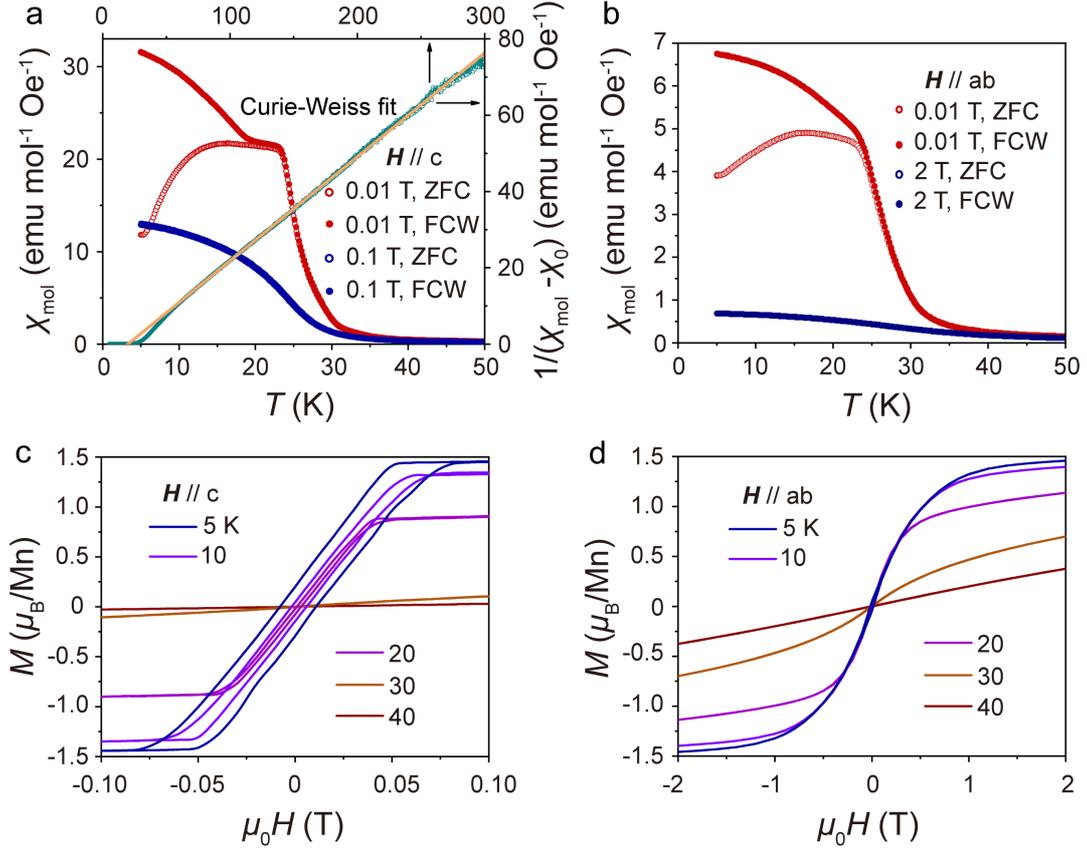

FIG. 3. DC magnetization of MnSb$_2$Te$_4$ single crystal. (a and b) Temperature dependence of magnetic susceptibility measured at different fields following zero-field-cooled (ZFC) and field-cooled (FC) processes, and inverse magnetic susceptibility as a function of temperature (top x axis and right y axis). The yellow line represents the Curie-Weiss fitting. Magnetic field is applied in out-of-plane (a) and in-plane (b) directions, respectively. (c and d) Field dependence of magnetization measured at different temperatures, and $\chi'$ and $\chi''$ correspond to the real and imaginary part of AC magnetic susceptibility. Magnetic field is out-of-plane (c) and in-plane (d), respectively.

Since DC magnetization data indicate the SG state of MnSb$_2$Te$_4$, the magnetic relaxation is investigated to further reveal magnetic properties of MnSb$_2$Te$_4$. Time-dependent magnetization curves in *H* // c are shown in Figs. 4(a)-4(b). In both ZFC and FC processes, magnetization continue to vary over minute-long periods. However, when the field is just increased to 0.1 T, magnetization in ZFC process shows no relaxation, demonstrating that MnSb$_2$Te$_4$ transforms into a field-induced FM state. Figures 4(c)-4(d) display the variation of the real ($\chi'$) and imaginary ($\chi''$) part of dynamic (AC) magnetic susceptibility at different frequencies as a function of temperature in *H* // c and *H* // ab, respectively. For *H* // c, the appearance of $\chi''$ are clear indication of the magnetic relaxation, and the peaks of $\chi'(T)$ and $\chi''(T)$ shift toward higher temperature as the frequency (*f*) increase, corresponding to the typical behavior of a SG state[39-41,52]. Figure S5a in the Supplemental Material displays the shift of $\chi''$ peaks toward lower temperatures at a larger AC driving field of 10 Oe with respect to those at an AC driving field of 5 Oe (see Fig. 3(c)), illustrating that the magnetic relaxation becomes faster at a larger AC driving field and the SG state is sensitive to the amplitude of AC driving field. As shown in Figs. S5b-S5d in the Supplemental Material, $\chi''$ is around zero and $\chi'$ becomes frequency-independent at a DC driving field higher than 600 Oe, indicating the disappearance of magnetic relaxation. As a result, the SG state in MnSb$_2$Te$_4$ can be turned into a field-induced FM state with a DC driving field up to 600 Oe as affirmed by peaks in $\chi'$, consistent with the DC magnetization results (see Figs. 2(a) and 3(a)). Furthermore, as shown in Fig. S6 in the Supplemental Material, by fitting the frequency dependence of the peak shift in $\chi'$ using $K = \Delta T_f/(T_f \Delta \log f)$, $K$ is calculated to be 0.005 at $H_{AC}$ = 5 Oe and 0.007 at $H_{AC}$ = 10 Oe in good agreement with the values (0.0045 ≤ $K$ ≤ 0.08) found in canonical SG systems[52]. For $H$ // ab, $\chi''(T)$ approximates zero and peaks of $\chi'(T)$ show no frequency-dependence, illustrating that magnetic relaxation does not exist in the in-plane direction. That is to say, the magnetization along ab plane is mainly from the field-bent magnetic moments, in good accordance with the DC magnetization results in Fig. 3.

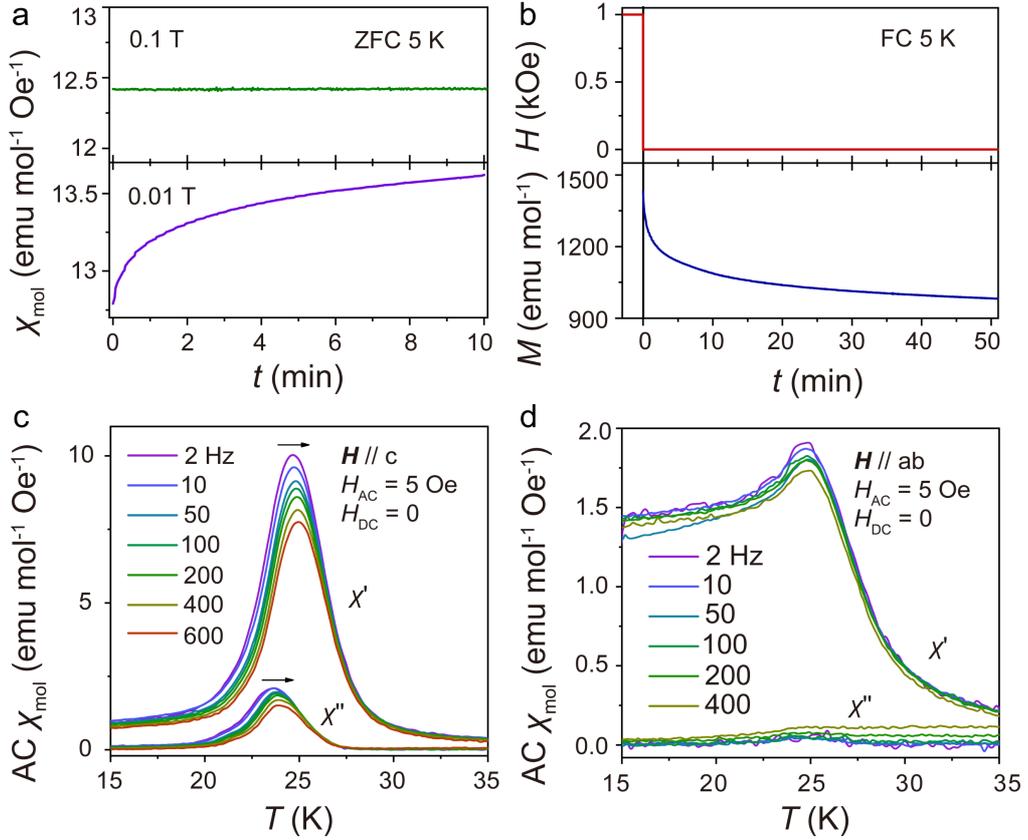

FIG. 4. Magnetic relaxation of MnSb$_2$Te$_4$ single crystal. (a) Time-dependent magnetic susceptibility at 5 K measured at different fields following a ZFC process. (b) Time-dependent magnetization at 5 K at zero field following a field-cooled (FC) process at a field of 1 kOe. (c and d) AC magnetic susceptibility measured at different frequencies. Magnetic field is out-of-plane (c) and in-plane (d), respectively.

The distribution of magnetic relaxation time ($\tau = 1/2\pi f$) is further investigated by the Cole-Cole plot for $\chi''$ and $\chi'$ (see Fig. S7 in the Supplemental Material) using a generalized Debye model (Experimental Section) to obtain the $\tau$-distribution parameter $\alpha$ (0 ≤ $\alpha$ ≤ 1). $\alpha$ is 0 for a single $\tau$, while $\alpha$ becomes greater for a broader distribution of $\tau$. The small derived $\alpha$ of MnSb$_2$Te$_4$ implies a rather simple magnetic relaxation process in the SG state. The $\alpha$ value (0.09, 0.41 at 20, 24 K, respectively) becomes greater with decreasing temperature, suggesting that the magnetic relaxation is largely affected by thermal energy. At a higher temperature, thermal activation could stimulate the transition between different magnetic states, generating a simpler magnetic relaxation process. When the temperature becomes lower, such transition needs to overcome higher energy barrier with a more complicated magnetic relaxation process. Thermal activation of the magnetic relaxation in SG state can be derived from the frequency dependent behavior of $\chi''$. As shown in Fig. S8 in the Supplemental Material, the data (black dots) of ln $(1/\tau)$ vs. $T^{-1}$ obtained at $H_{DC}$ = 0, $H_{AC}$ = 5 and 10 Oe follow the Arrhenius law, ln $(1/\tau)$ = $-E_a/k_BT$ + ln $(1/\tau_0)$, where $\tau$ is the relaxation time, and $T$ is derived from peaks in $\chi''$[39-41]. Accrodingly, the effective thermal activation energy is calculated to be 579 meV at $H_{AC}$ = 5 Oe, and 324 meV at $H_{AC}$ = 10 Oe, suggesting that the faster and easier magnetic relaxation in SG state with a larger AC driving field.

Different from the AFM nature of MnBi$_2$Te$_4$[15,29], MnSb$_2$Te$_4$ exhibits a SG state behavior below $T_{SG}$ ~ 24 K, which results from the much more pronounced cation intermixing in MnSb$_2$Te$_4$ with regard to that in MnBi$_2$Te$_4$. The cation intermixing as high as 28.9(7)% Sb on the Mn (3a) site and 19.3(6)% Mn on the Sb (6c) site may significantly weaken the Mn spin correlation that weakens the antiferromagnetism. At low temperature, MnSb$_2$Te$_4$ can maintain considerable magnetic moment with no applied field as illustrated in Figs. 3(c) and 4(b). The interplay between the intrinsic magnetism and topology arising from Sb-Te layer may lead to the realization of zero-magnetic-field QAH effect in MnSb$_2$Te$_4$, breaking no odd-layer restraint like the AFM-MB$_2$T$_4$[14].

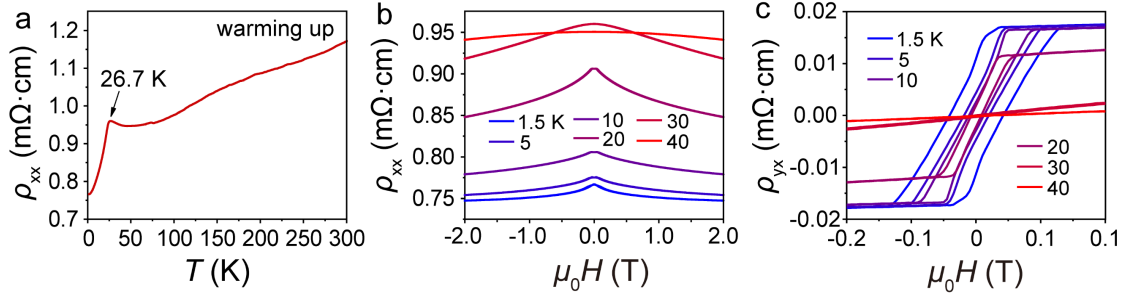

FIG. 5. Transport properties of MnSb$_2$Te$_4$ single crystal. (a) Temperature dependence of longitudinal resistivity $\rho_{xx}$ measured from 1.5 K to room temperature. (b) Magnetoresistivity (MR) curves in an out-of-plane magnetic field at varied temperatures from 1.5 K to 40 K. (c) Magnetic field dependence of Hall resistivity traces measured at the same condition as (b).

Furthermore, electrical transport measurements were carried out on a MnSb$_2$Te$_4$ single crystal with a thickness of ~ 35 μm to probe the transport property. Figure 5(a) shows displays the metallic temperature dependence of longitudinal resistivity $\rho_{xx}$ in the absence of an external magnetic field from 1.5 K to room temperature. At high temperature, $\rho_{xx}$ rises with increasing temperature, showing a metallic behavior. At low temperature, an upturn of $\rho_{xx}$ takes place at ~ 26.7 K, which can be attributed to the change in spin-flip scattering of charge carriers. Figure 5(b) displays the magnetoresistivity (MR) curves measured in **H**//c at varied temperatures. The drop of MR with increasing magnetic field is indicative of the alignment of magnetic domains. In the presence of magnetic field, the localized magnetic moments become more aligned, reducing the scattering of charge carriers and leading to the drop of MR. Figure 5(c) presents the magnetic field dependence of Hall resistivity measured in **H**//c at varied temperatures. The positive slopes of the Hall curves demonstrate a p-type conductor with a hole density of 1.8×10$^{20}$ cm$^{-3}$, and the mobilityof 45 cm$^2$ V$^-$

$^{-1}\cdot s^{-1}$ at 1.5 K. The high carrier concentration is probably caused by the cation anti defects and Te vacancies. An anomalous Hall effect featured a hysteresis loop is observed, consistent with the magnetic behavior in *M-H* curves (see Fig. 3(c)) and recent reports on few-layer $MnSb_2Te_4$[56]. Furthermore, the demagnetization steps the Hall curves at 1.5 K is a signal of relatively complicated local magnetic structures transforming from one to another, which is distinct from those of FM metals, again confirming the SG state[57-59]. The demagnetization steps disappear above 5 K, implying that the thermal energy could overcome the energy barrier of the transforming process, which is in good accordance with the magnetic measurements.

## IV. Conclusions

We have developed growth route for $MnSb_2Te_4$ single crystal, allowing the experimental exploration of a new member of $MnBi_2Te_4$-family material[29]. Similar to $MnBi_2Te_4$, the growth of high-quality $MnSb_2Te_4$ single crystals requires prolonged annealing in a narrow temperature range near 893 K due to its thermal metastability. The crystal structure analysis based on SCXRD data revealed the pronounced cation intermixing between Mn and Sb and lead to an unexpected SG state below $T_{SG}$ ~ 24 K, accompanied with strong magnetic anisotropy with easy axis along the c direction. This magnetic behavior makes $MnSb_2Te_4$ quite different from $MnBi_2Te_4$ with a well-established AFM state. Instead, it presents a frustrated spin system owing to the FM-AFM competitions. Furthermore, $MnSb_2Te_4$ displays considerable magnetic moment at zero applied field, as evidenced in the hysteresis loop of *M-H* and Hall curves in *H* // c. The interaction of net magnetism and possible nontrivial topology in the SG-type $MnSb_2Te_4$ enables an attractive testbed for the discovery and realization of exotic quantum phenomena without the restraint of layer number in the AFM-type $MnBi_2Te_4$-family materials.

## Acknowledgements

The authors acknowledge the financial support by the National Natural Science Foundation of China (Grant No. 51991340, 21975140) and the Ministry of Science and Technology of China (Grants No. 2018YFA0307100, and No. 2018YFA0305603). This work at Rutgers is supported by the Beckman Young Investigator award.